\documentstyle[prl,epsf,aps,floats]{revtex}
\begin{document}
\draft
\twocolumn[\hsize\textwidth\columnwidth\hsize\csname @twocolumnfalse\endcsname
\title{Mobile small polaron}

\author{A.S. Alexandrov$^{1}$ and P.E. Kornilovitch $^{2}$}

\address{$^{1}$ Department of Physics, Loughborough University,
Loughborough LE11 3TU, United Kingdom \\
$^{2}$ Blackett Laboratory, Imperial College, London SW7 2BZ, 
                       United Kingdom }

\date{\today}

\maketitle

\begin{abstract}

Extending the Fr\"ohlich polaron problem to a {\em discrete}
ionic lattice we study a  polaronic state with a small radius
of the wave function but a large size of the lattice distortion.
We calculate the energy dispersion and the effective mass 
of the polaron with the $1/\lambda$ perturbation theory and 
with the exact Monte Carlo method in the nonadiabatic and 
adiabatic regimes, respectively. The ``small'' Fr\"ohlich polaron 
is found to be lighter than the small Holstein polaron by one 
or more orders of magnitude.

\end{abstract}


\vskip2pc]
\narrowtext

A free electron interacting with the dielectric polarisable continuum
was studied by Pekar \cite{pek} and Fr\"ohlich \cite{fro}
in the strong and weak coupling limit, respectively. This is
the case of carriers interacting with optical phonons in
ionic crystals under the condition that the size of the self-trapped
state is large compared to the lattice constant so the lattice
discreteness is irrelevant \cite{mit}. The most sophisticated treatment 
of this ``{\em large}'' or ``continuum'' polaron is due to Feynman and
co-workers \cite{fey} with the path-integral method,
substantially extended in the past decade \cite{dev}. This treatment
leads to a mass enhancement, but not to a hopping conduction or to a
narrow polaron band.

When the electron-phonon coupling constant $\lambda$ is large,
all the states in the Brillouin zone are involved in the formation
of the polaron wave function, so the polaron radius becomes comparable
with the lattice constant $a$ and the continuum approximation is no
longer valid. Basic features of the {\em small} polaron were well
recognised a long time ago by Tjablikov \cite{tja}, Yamashita and
Kurosawa \cite{yam}, Sewell \cite{sew}, Holstein \cite{hol},
Lang and Firsov \cite{lan} and others, and are described
in several review papers and textbooks \cite{app,fir,bry,mah,alemot}.
So far, analytical and numerical studies have been mainly
confined to the Holstein model with a short-range electron-phonon
interaction. Exact diagonalization of several vibrating molecules
coupled with one electron \cite{alekab,feh}, variational \cite{rom,lam} 
and Monte Carlo calculations \cite{rad} revealed an excellent 
agreement with analytical results of Holstein \cite{hol} and 
Lang and Firsov \cite{lan} for the energy of the ground
state and first excited states at large $\lambda$.
Polaron mass is very large in the Holstein model, unless
phonon frequencies are extremely high. The size of the region,
where  the small Holstein polaron is localised, is about the same
as the size of the lattice distortion, each of the order of the
lattice constant. Both sizes are almost identical also
for the large Fr\"ohlich polaron, but much larger.

In this Letter we study a  problem of the lattice polaron
with a long-range Fr\"ohlich interaction \cite{fro2}.
This quasiparticle has a small (atomic) size of the electron 
localization region but a large size  of the lattice distortion. 
While the large Fr\"ohlich polaron is heavier than the large 
Holstein polaron, the {\em small} Fr\"ohlich polaron (SFP) 
turns out to be much lighter than the small Holstein polaron 
(SHP) with the same binding energy. We argue that SFPs are 
relevant quasiparticles in the cuprates.

A quite general electron-phonon lattice Hamiltonian
with one electron and the ``density-displacement'' type of
interaction is given by \cite{hol,fir,alemot}
\begin{eqnarray}
H = & - & \sum_{\bf n n'} t_{\bf n n'} c^{\dagger}_{\bf n'} c_{\bf n}
+ \sum_{{\bf q} \alpha} \hbar \omega_{{\bf q} \alpha}
( d^{\dagger}_{{\bf q} \alpha} d_{{\bf q} \alpha} + 1/2)  \cr
  & - & \sum_{{\bf mn} \alpha} f_{{\bf m} \alpha}({\bf n})
c^{\dagger}_{\bf n} c_{\bf n} \xi_{{\bf m} \alpha}  .
\label{zero}
\end{eqnarray}
Here $\alpha$ corresponds to the different phonon modes,
$\xi_{{\bf m} \alpha}$ is a normal coordinate at
 site ${\bf m}$, and $f_{{\bf m} \alpha}({\bf n})$ is
the {\em force} between the electron at site ${\bf n}$ and
the normal coordinate $\xi_{{\bf m} \alpha}$.

If characteristic phonon frequencies are large compared to
the electron kinetic energy, $\hbar \omega > t$, (nonadiabatic
regime) then one can apply a powerful analytic method,
 based on the Lang-Firsov canonical
transformation \cite{lan} and subsequent
 $1/\lambda$ perturbation technique.
Introducing the phonon operators  as $\xi_{{\bf m} \alpha} =
\sum_{\bf q} ( u_{{\bf m q} \alpha} d^{\dagger}_{{\bf q} \alpha} +
u^{\ast}_{{\bf m q} \alpha} d_{{\bf q} \alpha} )$
with $u_{{\bf m q} \alpha} = \hbar^{1/2}
(2 N M \omega_{{\bf q} \alpha} )^{-1/2}
e^{i {\bf q m} }$, $N$ the number of sites, and $M$
the ion mass, one obtains the transformed Hamiltonian 
\begin{eqnarray}
\tilde{H}&=& e^{-S} H e^{S}
= - \sum_{{\bf n'} \neq {\bf n}} \hat{\sigma}_{\bf n' n}
c^{\dagger}_{\bf n'}c_{\bf n} \cr
&-&E_{p}\sum_{\bf n}c_{\bf n}^{\dagger} c_{\bf n}
+ \sum_{{\bf q} \alpha} \omega_{{\bf q} \alpha}
(d_{{\bf q} \alpha}^{\dagger} d_{{\bf q} \alpha}+1/2) .
\label{one}
\end{eqnarray}
Here  $S = \sum_{{\bf m n q} \alpha}
(\hbar \omega_{{\bf q} \alpha})^{-1} u_{{\bf m q} \alpha}
f_{{\bf m} \alpha}({\bf n}) c^{\dagger}_{\bf n} c_{\bf n}
d^{\dagger}_{{\bf q} \alpha} - h.c. $, and $E_p$ is the familiar polaronic
shift,
\begin{equation}
E_p = \sum_{{\bf m m' q} \alpha}
\frac{1}{2N M \omega^2_{{\bf q} \alpha}}
f_{{\bf m} \alpha}( 0 ) f_{{\bf m'} \alpha}( 0 )
\cos{ {\bf q ( m - m') }}  .
\label{two}
\end{equation}
The polaronic shift is the natural measure of the strength of
the electron-phonon interaction. It defines
the electron-phonon coupling constant as $\lambda = E_p/zt$,
where $z$ is the lattice coordination number.
The first term in Eq. (\ref{one}) contains the transformed
hopping integral $\hat \sigma_{\bf n n'}$, which depends
on the phonon operators as
\begin{displaymath}
\hat{\sigma}_{\bf n n'}=t_{\bf n n'} \exp
\left[ \sum_{{\bf m q} \alpha}
\frac{ f_{{\bf m} \alpha} ({\bf n}) - f_{{\bf m} \alpha} ({\bf n'}) }
{\hbar \omega_{{\bf q} \alpha} }  \right.  \makebox[2.cm]{}
\end{displaymath}
\vspace{-0.7cm}
\begin{equation}
\left. \makebox[2.cm]{}
\times ( u_{{\bf m q} \alpha} d^{\dagger}_{{\bf q} \alpha}  -
  u^{\ast}_{{\bf m q} \alpha} d_{{\bf q} \alpha} )   \right] .
\label{three}
\end{equation}
At large $\lambda$ the hopping term in Eq. (\ref{one}) can
be treated as a perturbation. Introducing a set of $N$ zero-order
Bloch eigenstates ( all with the same energy $-E_p$ )
$ |{\bf k},0 \rangle =N^{-1/2}\sum_{\bf n}
c^{\dagger}_{\bf n} \exp (i {\bf k \cdot n})|0 \rangle $,
one readily  calculates the lowest energy levels in a crystal.
Up to the second order in the hopping integral, the result is
\begin{eqnarray}
E({\bf k})
 & = & - E_{p} - \sum_{{\bf n}\neq 0}
t_{{\bf n} 0} \, e^{ - g^2({\bf n}) } \exp (-i {\bf k \cdot n}) \cr
 & - & \sum_{ {\bf k'},n_{{\bf q} \alpha} }
{|\langle {\bf k},0 | \sum_{\bf n n'} \hat{\sigma}_{\bf n n'}
c^{\dagger}_{\bf n'} c_{\bf n}|{\bf k'},
n_{{\bf q} \alpha} \rangle |^{2} \over{\hbar \sum_{{\bf q} \alpha}
\omega_{{\bf q} \alpha} n_{{\bf q} \alpha}}}.
\label{five}
\end{eqnarray}
Here $|{\bf k'},n_{{\bf q} \alpha} \rangle$ is an excited
state of the unperturbed Hamiltonian with one electron and
at least one phonon, $n_{{\bf q} \alpha}$ is the phonon
occupation number. The second term in Eq. (\ref{five}), which
is linear with respect to the bare hopping $t_{\bf n n'}$,
determines the dispersion of the polaron band 
with a band-narrowing exponent (at zero temperature)
\begin{equation}
g^{2}({\bf n})  = \sum_{{\bf q} \alpha}
\frac{1}{2 N M \hbar \omega^3_{{\bf q} \alpha}}
\times \makebox[3.cm]{}
\label{seven}
\end{equation}
\vspace{-0.6cm}
\begin{displaymath}
\sum_{\bf m m'} \left[  f_{{\bf m} \alpha}(0) f_{{\bf m'} \alpha} (0)
- f_{{\bf m} \alpha}(0) f_{{\bf m'} \alpha} ({\bf n}) \right]
\cos {\bf q(m-m')} .
\end{displaymath}
The third term in Eq.(\ref{five}), quadratic in $t_{\bf n n'}$,
yields a negative almost ${\bf k}$-{\em independent}
correction of the order of $1/\lambda^{2}$ to the polaron level shift.
It is unrelated to the polaron effective mass and the polaron
tunneling mobility.

In general, there is no simple relation between the polaronic
shift $E_p$ and the exponent $g^2$ which describes the mass
enhancement, as one can see from Eq.~(\ref{two}) and Eq.~(\ref{seven}).
We now consider the case of a single dispersionless phonon mode
$\omega_{{\bf q} \alpha} = \omega$ and the nearest-neighbor
hopping with an amplitude $t$. One obtains
\begin{equation}
E_p = \frac{1}{2 M \omega^2} \sum_{\bf m} f^2_{\bf m} ( 0 ) ,
\label{eight}
\end{equation}
\vspace{-0.5cm}
\begin{equation}
g^2 \equiv g^2(1) = \frac{1}{2 M \hbar \omega^3} \sum_{\bf m}
\left[ f^2_{\bf m} ( 0 ) - f_{\bf m} (0) f_{\bf m} (1) \right].
\label{nine}
\end{equation}
The effective mass renormalisation is $m^{\ast}/m= e^{g^2}$, where $m$
is the bare band mass and $1/m^{\ast} = \partial^{2} E({\bf k})/
\partial (\hbar k)^{2}$ with $k \rightarrow 0$.
If the interaction is local,
$f_{\bf m}({\bf n}) = \kappa \delta_{\bf mn}$ (Holstein model),
then $g^2 = E_p/(\hbar \omega)$. In general, one has
$g^2 = \gamma E_p/(\hbar \omega)$ with a numerical coefficient
$\gamma = 1 - \sum_{\bf m} f_{\bf m} (0) f_{\bf m} (1) /
\sum_{\bf m'} f^2_{\bf m'} ( 0 ) $, which is less than unity for the
canonical Fr\"ohlich interaction \cite{ale}.

\begin{figure}[t]
\begin{picture}(200,120)(20,20)
\multiput(80,100)(30,0){5}{\circle{6}}
\multiput(77,70)(30,0){5}{$\times$}
\put(166,90){{\bf m}}
\put(108,63){{\bf n}}
\put(113,73){\vector(1,0){24}}
\put(123,77){$t$}
\end{picture}
\caption{
One-dimensional model of the small Fr\"ohlich polaron
on chain ($\times$) interacting with all ions of chain ($\circ$).
}
\label{fig1}
\end{figure}
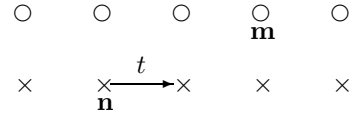

To calculate $\gamma$ explicitly we introduce one and two-dimensional
lattice models with a long-range Coulomb interaction between
an electron and ions (see Fig.1). The electron in a Wannier state
on a site ${\bf n}$ of the infinite chain (plane) ($\times$) interacts
with the vibrations of {\em all} ions of another chain (plane) ($\circ$)
polarised in the direction perpendicular to the chains.
A strong  coupling of carriers with  $c$-axis polarised
phonons  ($\hbar \omega \simeq 75$ meV) has been established
experimentally  in YBa$_2$Cu$_3$O$_{6+x}$ \cite{tim}. Because of a
low $c$-axis conductivity and high phonon frequency, this coupling
is not screened representing an example of a long-range
Fr\"ohlich interaction. In this way our model mimics
a hole on the CuO$_2$ plane (chain $\times$) coupled with the
c-axis apical oxygen vibrations (chain $\circ$) in the cuprates.
The corresponding force is given by
\begin{equation}
f_{\bf m} ({\bf n}) =  \frac{\kappa}
{( | {\bf m} - {\bf n} |^2 + 1)^{3/2}} .
\label{ten}
\end{equation}
Here the distance along the chains $| {\bf m} - {\bf n} |$ is measured
in lattice constants $a$, and the inter-chain distance is also $a=1$.
For this long-range interaction, one obtains
$E_p = 1.27 \kappa^2/(2M\omega^2)$,
$g^2 = 0.49 \kappa^2/(2M\hbar\omega^3)$, and
$g^2 = 0.39 E_p/(\hbar \omega)$.
The effective mass renormalisation is much smaller than in the Holstein
model, roughly as $m^{\ast}_{SFP} \propto \sqrt{m^{\ast}_{SHP}}$.

Our analytical consideration is applied if $\omega \geq t$, and
$\lambda \gg 1$. To extend the results to the adiabatic case and to the
intermediate coupling we apply a continuous-time path-integral Quantum
Monte Carlo (QMC) algorithm, developed recently \cite{kor}.
This method is free from any systematic finite-size, finite-time-step
and finite-temperature errors and allows for exact 
calculation of the ground-state energy and the effective
mass of the lattice polaron for any electron-phonon interaction.
The method was tested on the one-dimensional (1D) Holstein model which
has been extensively studied by other methods. Excellent agreement
with exact diagonalization \cite{alekab,feh}, density-matrix
renormalization group \cite{jec} and variational \cite{rom}
results was found for both the ground-state energy and
effective mass.

\begin{figure}[t]
\begin{center}
\leavevmode
\hbox{
\epsfxsize=8.6cm
\epsffile{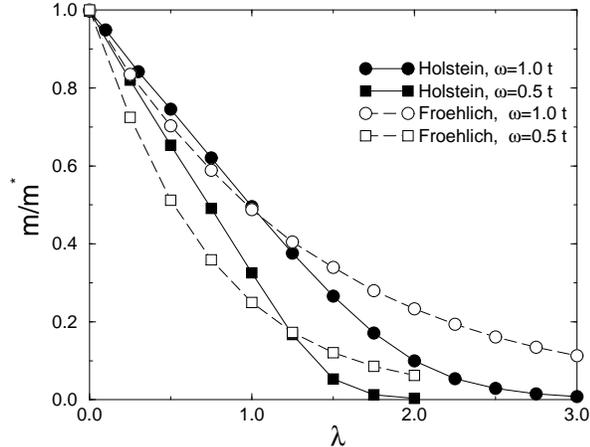}
}
\end{center}
\vspace{-0.5cm}
\caption{
Inverse effective polaron mass in units of $1/m = 2ta^2/\hbar^2$
for the one-dimensional Holstein and Fr\"ohlich [Eq. (\ref{ten})]
models. Circles: $\omega = 1.0 \, t$; squares:
$\omega = 0.5 \,t $.
}
\label{fig2}
\end{figure}
\begin{figure}[t]
\begin{center}
\leavevmode
\hbox{
\epsfxsize=8.6cm
\epsffile{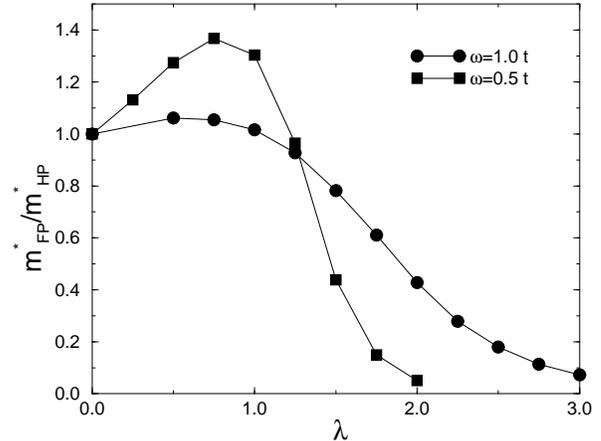}
}
\end{center}
\vspace{-0.5cm}
\caption{
The ratio of the effective masses of the Fr\"ohlich and Holstein polarons
in 1D. Fr\"ohlich polaron is heavier at small $\lambda < 1.25$ but much
lighter at $\lambda>1.25$.
}
\label{fig3}
\end{figure}

Exact polaron masses of the one-dimensional model defined by
Eq.~(\ref{ten}) and Fig.~\ref{fig1}, are compared with 1D Holstein
polaron masses in Fig.~\ref{fig2}. For both phonon frequencies
$\hbar \omega = 1.0\, t$ and $0.5\, t$ we found SFP to be
{\em heavier} than SHP at small $\lambda < 1$, but {\em much}
lighter than SHP in the strong-coupling regime $\lambda > 1.5$.
The mass ratio reaches 1 order of magnitude at $\lambda = 2.75$
for $\hbar \omega = 1.0 \, t$ and at $\lambda = 1.75$ for
$\hbar \omega = 0.5 \, t$. This is in accordance with our analytical
approach in the $\hbar \omega > t$ regime. Thus the mass
ratio $m^{\ast}_{FP}/m^{\ast}_{HP}$ is a non-monotonic
function of $\lambda$ (see Fig.~\ref{fig3}). This is a
consequence of the fact that $m^{\ast}_{FP}(\lambda)$
is well fitted by a single exponential function,
$\exp(0.73 \lambda)$ for $\hbar \omega = 1.0 \, t$
and  $\exp(1.40 \lambda)$ for $\hbar \omega = 0.5 \, t$.
This is not so for the Holstein polaron, in which case
a crossover between two regimes occurs at $\lambda \sim 1.5$.
It is interesting that the numerical exponents found are only
slightly smaller than that follow from the Lang-Firsov transformation,
$\exp(0.78\lambda)$ and $\exp(1.56\lambda)$ respectively.
This shows the excellent accuracy of this transformation even in
the intermediate region of parameters, $\lambda \sim 1$ and $\hbar
\omega/t \sim 1$. Note, however, that the exact exponent deviates
more and more from the Lang-Firsov approximation with a decreasing
adiabatic ratio $\hbar \omega/t$. This is in agreement with the
exact diagonalization of a two-site model \cite{alekab}, where it
was shown that the Lang-Firsov approximation overestimates
the polaron mass in the adiabatic regime.

We also compared our exact QMC masses with the canonical weak-
\cite{fro} and strong-coupling \cite{pek} continuum polaron theory,
where the bandwidth is assumed to be infinite. To make such a
comparison meaningful we determine the Fr\"ohlich coupling constant
$\alpha$ in such a way, that the ground-state energy $E_{0}$ of the continuum
approximation is the same as the one in our model. Then we calculate
the continuum-case mass and compare with our $m^{\ast}_{SFP}(\lambda)$.
In the Fr\"ohlich weak-coupling regime one has
$E_{0}=-\alpha \, \hbar\omega$ and $m^{\ast}_{c}=(1+\alpha/6)\,m$.
This mass appears to be well below our $m^{\ast}_{FP}$ for $\lambda <1$.
For instance, for $\lambda=0.5$ and $\hbar\omega =t$, the continuum mass is
$m^{\ast}_c=1.119\,m$ while our result is $m^{\ast}_{FP}=1.422\,m$.
However, in the strong-coupling regime, $\lambda>1$, the
continuum approximation overestimates the mass.
Using Pekar's ground state energy, $E_{0}=-0.1085\alpha^2 \, \hbar\omega$,
and mass $m^{\ast}_{c}=0.021\alpha^4 \, m$ for $\lambda=2$ and $\hbar\omega=t$
we find $m^{\ast}_c = 17.2\,m$ which is much larger than our mass,
$m^{\ast}_{FP}=4.29\,m$. This difference does not depend very much on
the dimensionality of the polaron. We notice also that if we take
into  account the intermediate coupling corrections to the ground
state energy of the strong-coupling Pekar polaron,
$E_{0}=-(0.109\alpha^2 + 2.836)\hbar\omega$ \cite{mia}, a continuum
polaron mass $m^{\ast}_{c}=1.074 \, m$ turns out to be much lighter than
the exact one for the same $\lambda$. These estimates underline the
crucial role of a finite bandwidth.

\begin{figure}[t]
\begin{center}
\leavevmode
\hbox{
\epsfxsize=8.6cm
\epsffile{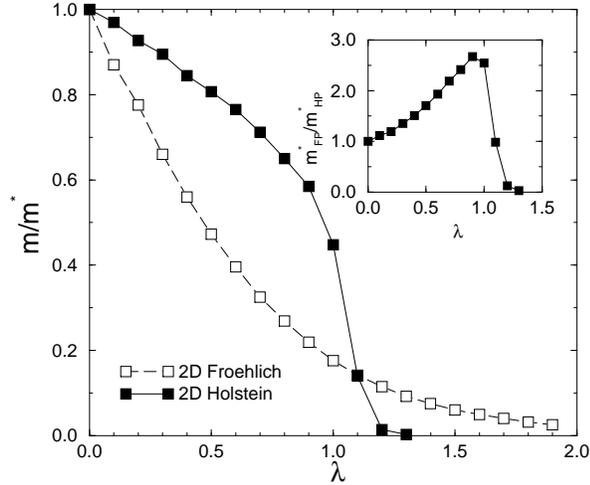}
}
\end{center}
\vspace{-0.5cm}
\caption{
Inverse effective polaron mass in units of $1/m = 2ta^2/\hbar^2$
for the two-dimensional Holstein and Fr\"ohlich [Eq. (\ref{ten})]
models. $\hbar\omega = 0.5 \, t$.
Inset: Ratio $m^{\ast}_{FP}/m^{\ast}_{HP}$.
}
\label{fig4}
\end{figure}

To check that the light small Fr\"ohlich polaron is not an artifact
of one dimension we calculated its mass for the {\em two}-dimensional (2D)
version of the model (\ref{ten}) and compared it with the 2D
Holstein polaron (see Fig.~\ref{fig4}). At $\lambda>1$ the mass ratio
$m^{\ast}_{SFP}/m^{\ast}_{SHP}$ (see inset) shows even more sharp
fall than in 1D. While SFP is 2.5 times heavier than SHP at $\lambda=1.0$,
they are equal at $\lambda=1.1$, and SFP is 36 times lighter at
$\lambda=1.3$ (at this coupling $m^{\ast}_{SHP}=400$ but 
$m^{\ast}_{SHP}=11$). The reason for such a
dramatic change is the {\em very} large mass of 2D SHP. At the
same time, the mass of SFP grows exponentially but smoothly, similar
to the 1D case. The best fit to QMC data is
$\exp (1.62\lambda+0.19 \lambda^2)$.

The physical reason for the  small mass of SFP lies in
the form of electron-phonon
interaction. A long-range interaction of the type of Eq.~(\ref{ten})
induces a lattice distortion which undergoes {\em less} changes when
the carrier hops to the neighboring site, than a distortion induced
by a short-range interaction. Namely, relative changes
are essential for the polaron mass. One should also emphasize
the new type of internal structure of SFP, which is best understood
in the extreme strong-coupling limit, $\lambda \rightarrow \infty$.
In this limit the Lang-Firsov transformation is exact, and the polaron
is localised on one site ${\bf n}$. Hence, the size of its wave function is
the atomic size. On the other hand, the lattice deformation, which
is proportional to the displacement force $f_{\bf m} ({\bf n})$,
spreads over a large distance. Its amplitude falls with the distance
as $|{\bf m-n}|^{-3}$ in our model. Thus we have a new situation when
the size of the polaron and the size of lattice deformation are
very different. Our findings suggest to generalise the definitions
of the polaron and polaron ``cloud'' to include this new possibility.

In conclusion, we have studied the small polaron problem with the
long-range Fr\"ohlich interaction. This polaron has
a small (atomic) size of the wave function but  a large size
of the lattice deformation.  The ``{\em small} Fr\"ohlich polaron''.
propagates in a narrow band with the effective mass much smaller
than that of the Holstein small polaron with the same binding energy.
We argue that small Fr\"ohlich (bi)polarons \cite{ale} as well as large
Fr\"ohlich (bi)polarons \cite{emi,sil,dev} are relevant quasiparticles 
in the cuprates, describing holes in the $CuO_{2}$
plane coupled with the lattice distortion by a long-range interaction.

We are grateful to J.\,T.\,Devreese, D.\,M.\,Eagles, H.\,Fehske, 
Yu.\,A.\,Firsov, W.\,M.\,C.\,Foulkes, V.\,V.\,Kabanov, 
E.\,K.\,Kudinov, and Guo-meng Zhao for illuminating discussions. 
PEK acknowledges the support by EPSRC, grant No. GR/L40113.


\begin{references}

\bibitem{pek}
S.I. Pekar,
Zh.\,Eksp.\,Teor.\,Fiz. {\bf 16}, 335 (1946).

\bibitem{fro}
H. Fr\"ohlich,
Adv.\,Phys. {\bf 3}, 325 (1954).

\bibitem{mit}
For an extensive review of the large polaron problem see 
T.\,K.\,Mitra {\em et al},
Phys.\,Rep. {\bf 153}, 91 (1987).

\bibitem{fey}
R.\,P.\,Feynman,
Phys.\,Rev. {\bf 97}, 660 (1955);
R.\,P.\,Feynman {\em et al},
Phys.\,Rev. {\bf 127}, 1004 (1962).

\bibitem{dev}
J.T.Devreese, in
{\em Encyclopedia of Applied Physics},
vol. 14, p. 383  (VCH Publishers, 1996).

\bibitem{tja}
S.\,V.\,Tjablikov,
Zh.\,Eksp.\,Teor,\,Fiz. {\bf 23}, 381 (1952).

\bibitem{yam}
J.\,Yamashita and T.\,Kurosawa,
J.\,Phys.\,Chem.\,Solids {\bf 5}, 34 (1958).

\bibitem{sew}
G.\,L.\,Sewell,
Phil.\,Mag. {\bf 3}, 1361 (1958).

\bibitem{hol}
T.\,Holstein,
Ann.\,Phys. {\bf 8}, 325-42; {\em ibid} p. 343 (1959).

\bibitem{lan}
I.\,G.\,Lang and Yu.\,A.\,Firsov,
Zh.\,Eksp.\,Teor.\,Fiz. {\bf 43}, 1843 (1962)
[ Sov.\,Phys.\,JETP {\bf 16}, 1301 (1963)].

\bibitem{app}
J.\,Appel, in
{\em Solid State Physics}, eds. F.\,Seitz, D.\,Turnbull, and
H.\,Ehrenreich, {\bf 21} ( Academic Press, 1968 ).

\bibitem{fir}
Yu.\,A.\,Firsov  (ed),
{\em Polarons}, ( Nauka, Moscow, 1975).

\bibitem{bry}
H.\,B\"ottger and V.\,V.\,Bryksin,
{\em Hopping Conduction in Solids},
( Academie-Verlag, Berlin, 1985).

\bibitem{mah}
G.\,D.\,Mahan,
{\em Many Particle Physics}, ( Plenum Press, 1990).

\bibitem{alemot}
A.\,S.\,Alexandrov and N.\,F.\,Mott,
{\em Polarons and Bipolarons},
(World Scientific, Singapore, 1995).

\bibitem{alekab}
A.\,S.\,Alexandrov, V.\,V.\,Kabanov, and D.\,K.\,Ray,
Phys.\,Rev.\,B {\bf 49}, 9915 (1994) and references therein.

\bibitem{feh}
H.\,Fehske, J.\,Loos, and G.\,Wellein,
Z.\,Phys.\,B {\bf 104}, 619 (1997) and references therein.

\bibitem{rom}
A.\,H.\,Romero, D.\,W.\,Brown, and K.\,Lindenberg,
J.\,Chem.\,Phys. {\bf 109}, 6540 (1998).

\bibitem{lam}
A.\,La\,Magna and R.\,Pucci,
Phys.\,Rev.\,B {\bf 55}, 14886 (1997).

\bibitem{rad}
H.\,De\,Raedt and A.\,Lagendijk,
Phys.\,Rev.\,B {\bf 27}, 6097 (1983).

\bibitem{fro2}
H.\,Fr\"ohlich,
Phys.\,Rev. {\bf 79}, 845 (1950).

\bibitem{ale}
For cuprates one estimates $\gamma \simeq 0.2$ 
[A.\,S.\,Alexandrov, Phys.\,Rev.\,B {\bf 53}, 2863 (1996); cond-mat/9807185]. 
Small polaron parameters for many oxides can be found in 
D.\,M.\,Eagles, Phys.\,Rev. {\bf 181}, 1278 (1969); 
J.\,Phys.\,C {\bf 17}, 637 (1984).

\bibitem{tim}
T.\,Timusk {\em et al}, in
{\em Anharmonic Properties of High-$T_{c}$ Cuprates},
eds. D.\,Mihailovi\'c {\em et al},
(World Scientific, Singapore, 1995), p.171.

\bibitem{kor}
P.\,E.\,Kornilovitch,
Phys.\,Rev.\,Lett {\bf 81}, 5382 (1998).

\bibitem{jec}
E.\,Jeckelmann and S.\,R.\,White,
Phys.\,Rev.\,B {\bf 57}, 6376 (1998).

\bibitem{mia}
S.\,J.\,Miyake,
J.\,Phys.\,Soc.\,Japan {\bf 41}, 747 (1976).

\bibitem{emi}
D.\,Emin, 
Phys.\,Rev.\,Lett. {\bf 62}, 1544 (1989).

\bibitem{sil}
S.\,Sil, A.\,K.\,Giri, and A.\,Chatterjee, 
Phys.\,Rev.\,B {\bf 43}, 12642 (1991)
.
\end{references}
\end{document}